\documentclass[runningheads]{llncs}
\usepackage{graphicx}
\usepackage{comment}
\usepackage{url}
\usepackage{xcolor}
\begin{document}
\title{Searching for Pneumothorax in Half a Million Chest X-Ray Images\thanks{Supported by Vector Institute Pathfinder Project}}
\author{Antonio Sze-To\inst{1}\and
Hamid Tizhoosh\inst{1,2}}
\authorrunning{To appear in AIME 2020 International Conference on AI in Medicine, USA}
\institute{KIMIA Lab, University of Waterloo, Waterloo, ON, Canda N2L 3G1
\email{hy2szeto@uwaterloo.ca}\\
\and
Vector Institute, Toronto, ON, Canada M5G 1M1\\
\email{hamid.tizhoosh@uwaterloo.ca}}
\maketitle              % typeset the header of the contribution
\begin{abstract}
Pneumothorax, a collapsed or dropped lung, is a fatal condition typically detected on a chest X-ray by an experienced radiologist. 
Due to shortage of such experts, automated detection systems based on deep neural networks have been developed. 
Nevertheless, applying such systems in practice remains a challenge. 
These systems, mostly compute a single probability as output, may not be enough for diagnosis. 
On the contrary, content-based medical image retrieval (CBIR) systems, such as image search, can assist clinicians for diagnostic purposes by enabling them to compare the case they are examining with previous (already diagnosed) cases.
However, there is a lack of study on such attempt.
In this study, we explored the use of image search to classify pneumothorax among chest X-ray images. 
All chest X-ray images were first tagged with deep pretrained features, which were obtained from existing deep learning models. 
Given a query chest X-ray image, the majority voting of the top K retrieved images was then used as a classifier, in which similar cases in the archive of past cases are provided besides the probability output. 
In our experiments, 551,383 chest X-ray images were obtained from three large recently released public datasets. Using 10-fold cross-validation, it is shown that image search on deep pretrained features achieved promising results compared to those obtained by traditional classifiers trained on the same features. 
To the best of knowledge, it is the first study to demonstrate that deep pretrained features can be used for CBIR of pneumothorax in half a million chest X-ray images. 

\keywords{Deep Learning  \and Chest X-ray Images \and Content-Based Image Retrieval (CBIR) \and Image Search \and Pneumothorax.}
\end{abstract}
\section{Introduction}
\begin{figure*}[htp]
  \centering
  %\fbox{\rule[-.5cm]{0cm}{4cm} \rule[-.5cm]{4cm}{0cm}}
  \includegraphics[width=0.9\textwidth]{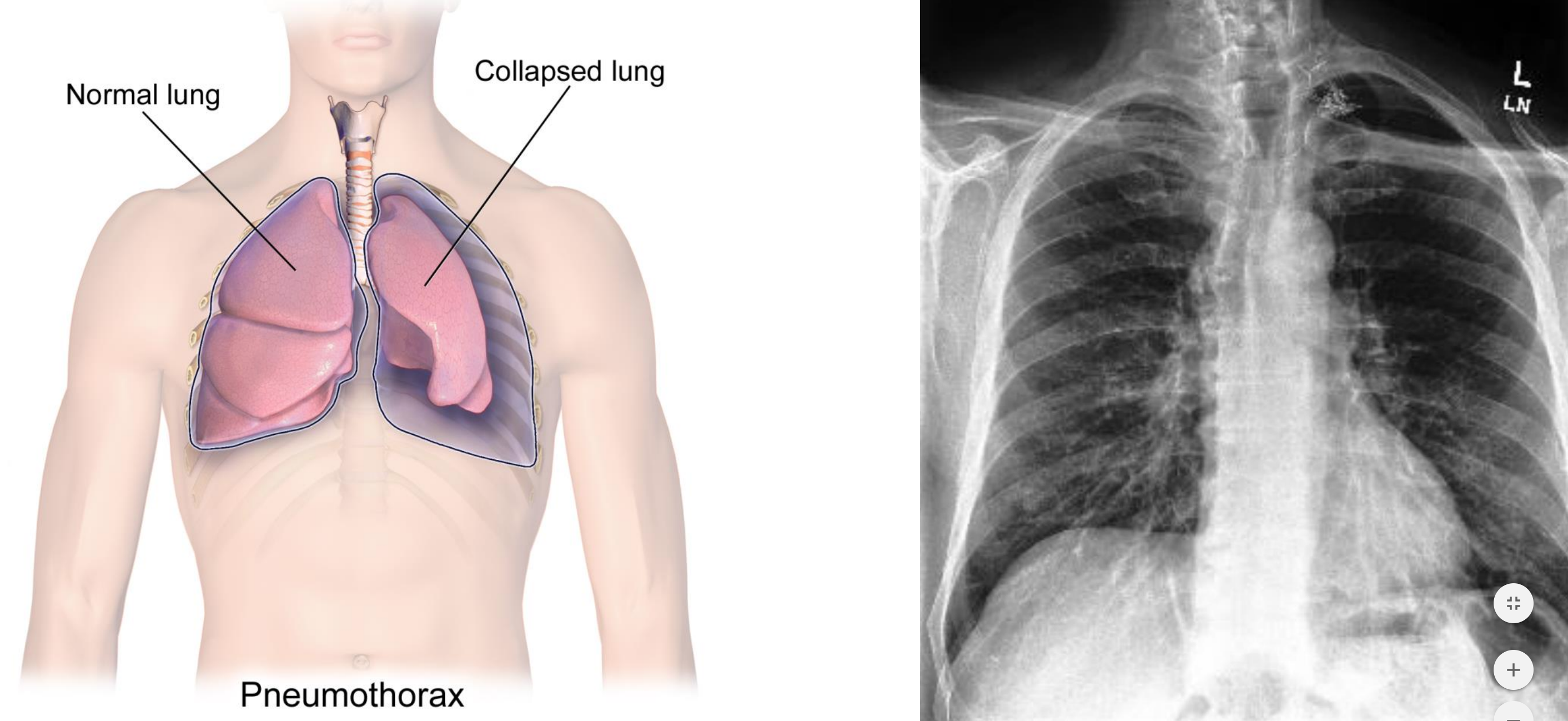}
  \caption{Left: A graphical illustration of pneumothorax \cite{richfield2014medical}. Right: A X-ray image of Pneumothorax in left lung visible thorough slight contrast different as a result of the lung collapse, obtained from CheXpert \cite{irvin2019chexpert}. %[Source of graphic: Medical gallery of Blausen Medical 2014. WikiJournal of Medicine 1 (2). DOI:10.15347/wjm/2014.010. ISSN 2002-4436.]
  }
\label{fig:demo}
\end{figure*}

Pneumothorax is a life-threatening emergency condition that can lead to death of the patient \cite{imran2017pneumothorax}.
It is an urgent situation \cite{zarogoulidis2014pneumothorax} where air enters the pleural space, i.e. the space between the lungs and the chest wall \cite{imran2017pneumothorax}.
An illustration of pneumothorax, and a sample chest X-ray image are provided in Fig \ref{fig:demo}.

%where (a) refers to a patient with no finding and (b) refers to a patient with pneumothorax.
Pneumothorax is typically detected on chest X-ray images by qualified radiologists \cite{zarogoulidis2014pneumothorax}. 
%It is, however, challenging, as pneumothorax has the characteristics of curved contour and smooth regions within dark region against the chest wall and ribs, and clavicles may overlap \cite{doi1997method}.
%Thus, pneumothorax may be missed or misclassified as other diseases easily \cite{doi1997method}.
As it is time-consuming and expensive to train qualified radiologists \cite{ker2017deep}, the supply of qualified radiologist is rather limited. 
%The large volume of work for radiologists on one hand and the low visibility of pneumothorax on the other hand have made the detection of pneumothorax quite difficult.  
%Thus, currently there are long worklists of chest X-ray images awaiting radiological review, understandably delaying time to diagnosis and treatment.
Since an incorrect or delayed diagnosis can cause harm to patients \cite{ker2017deep}, it is vital to develop computer-aided approaches to assist radiologists.

%Pneumothorax (collapsed lung or dropped lung) \cite{zarogoulidis2014pneumothorax,imran2017pneumothorax} refers to the entry of air into the pleural space, i.e. the space between the lungs and chest wall \cite{imran2017pneumothorax}.
%It is an urgent situation \cite{zarogoulidis2014pneumothorax} and can be life-threatening emergency \cite{imran2017pneumothorax}.
%Pneumothorax is mostly diagnosed by chest X-ray images \cite{imran2017pneumothorax}.
%Fig. \ref{fig:problem} provides a graphical illustration of Pneumothorax, where (a) refers to a patient with no finding and (b) refers to a patient with Pneumothorax.
%Detecting Pneumothorax on chest X-ray images is challenging, as expertise of radiologists is required \cite{ker2017deep}.
%It is time-consuming and expensive to train a qualified radiologist.
%As a delayed diagnosis can cause harm to patients \cite{ker2017deep}, it is important for the development of computer-aided detection approaches. 

Due to its recent success, an increasing number of studies have adopted deep learning or deep neural networks (DNNs) to detect pneumothorax or other thoracic diseases in chest X-ray images \cite{wang2017chestx,rajpurkar2017chexnet,irvin2019chexpert}.
%In essence, deep learning refers to the use of Deep Neural Networks (DNN), defined as artificial neuron networks with at least 3 hidden layers \cite{lecun2015deep}).
%Its success is built upon the advent of fast graphics processing units (GPU), and the availability of large amount of labeled data \cite{lecun2015deep}. 
%It has been shown to outperform other machine-learning techniques in a variety of tasks \cite{lecun2015deep}, such as beating the world champion in GO \cite{silver2016mastering}.
Since the availability of ChestXray8 (or its later version ChestXray14) \cite{wang2017chestx}, one of the largest publicly available chest X-ray datasets, it is possible to train DNNs as classifiers to output a probability for certain thoracic diseases.
Its performance has been reported to achieve or exceed the level of qualified radiologists on certain diseases such as pneumonia \cite{rajpurkar2017chexnet}.
However, a single probability output may not be enough for convincing diagnosis.  

As an alternative, image search not only provides a probabilistic output but also similar cases from the past cases. %, allowing for a comparison of patient histories and treatment profiles \cite{zhang2014towards}, hence representing a virtual ``second opinion'' for diagnostic purposes. 
%Retrieving similar images given a query image is known as Content-Based Image Retrieval (CBIR) \cite{zhou2017recent} or Content-Based Medical Image Retrieval (CBMIR) \cite{das2017overview} for medical applications.
Retrieving similar images given a query image for medical applications is an application of Content-based Image Retrieval (CBIR) \cite{tzelepi2018deep} for medical images. 
It is also known as Content-Based Medical Image Retrieval (CBMIR) \cite{das2017overview}.
CBMIR can help doctors in retrieving similar images and case histories for understanding the specific patient’s disease or injury status and can also help to exploit the information in corresponding medical reports \cite{das2017overview}. 
It may also help radiologists in preparing the report for particular diagnosis \cite{das2017overview}. 
%Classification of X-ray images and searching in archives of radiology have been interesting research tasks \cite{camlica2015medical,tizhoosh2015barcode}. 
While deep learning methods have been applied to image retrieval tasks in recent studies \cite{tzelepi2018deep}, there is less attention on exploring deep learning methods for CBMIR tasks \cite{qayyum2017medical}.

In this study, we explored the use of image search, based on features obtained from DNNs i.e. deep features, to detect pneumothorax among more than 550,000 chest X-ray images obtained from three large recently released labelled datasets, namely ChestX-ray14 \cite{wang2017chestx}, CheXpert \cite{irvin2019chexpert} and MIMIC-CXR \cite{goldberger2000physiobank,johnson2019mimic}.
In our experiments, all chest X-ray images were first tagged with DenseNet121 deep features \cite{huang2017densely}.
Given a query chest X-ray image, the majority voting of the top $K$ retrieved  X-ray images was then used as a classifier.
%We proposed a re-ranking algorithm to sort the output of search. 
%Under 10-fold cross-validation, it is shown that the use of image search as a classifier is comparable to traditional classifiers trained with deep features in Pneumothorax detection.
%To our knowledge, it is the first study on searching Pneumothorax in more than half a million chest X-ray images.

%-------------------------------------------------------------------------
%-------------------------------------------------------------------------

\section{Methodology}
\begin{figure*}[htp]
  \centering
  %\fbox{\rule[-.5cm]{0cm}{4cm} \rule[-.5cm]{4cm}{0cm}}
  \includegraphics[width=1.0\textwidth]{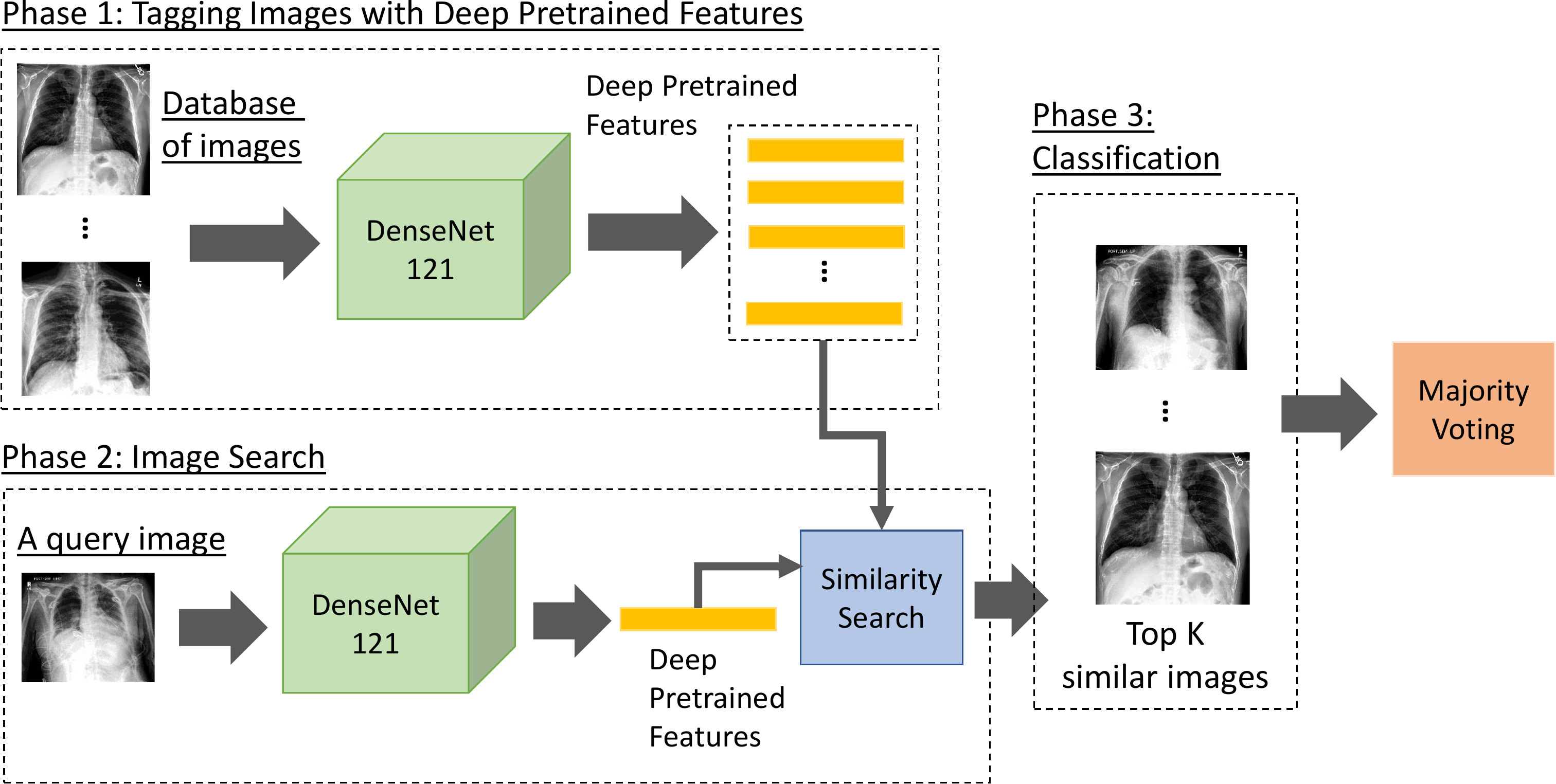}
  \caption{
An overview of using image search as a classifier to detect pneumothorax in chest X-ray images.
The method is composed of three phases. Phase 1: Tagging images with Deep Pretrained Features. All images in the database are tagged with deep pretrained features. Following \cite{rajpurkar2017chexnet}, DenseNet121 \cite{huang2017densely}, pretrained on ImageNet \cite{russakovsky2015imagenet} dataset, is adopted among existing models for converting a chest X-ray image into a feature vector with 1024 dimensions. Phase 2: Image Search. The query image is first tagged with deep pretrained features. Then, the distance between the query features and all other features in the database are computed to find the most similar images. Phase 3: Classification. The majority voting of the retrieved images is used as a classifier.
}
\label{fig:method}
\end{figure*}

The proposed method of using image search as a classifier comprises of three phases (Fig. \ref{fig:method}): 1) Tagging images with features (all images in the database are tagged with deep pretrained features), 2) Receiving a query image (tagging with features and calculating its distance with all other features in the database to find the most similar images), and 3) Classification (majority voting among the labels of retrieved images).

\textbf{Phase 1: Tagging Images with Deep Pretrained Features} -- In this phase, all chest X-ray images in the database are tagged with deep pretrained features. 
To represent a chest X-ray image as a feature vector with a fixed dimension, the output of the fully-connected layer before the classification layer of a deep convolutional neuron network (DCNN) is used. 
These values are denoted as deep features \cite{zhang2018unreasonable}, or technically deep pretrained features if the DCNN is pretrained with other datasets. 
In other words, the DCNN is considered as a feature extractor to convert a chest X-ray image into an $n$-dimensional feature vector.
In this study, following \cite{rajpurkar2017chexnet}, DenseNet121 \cite{huang2017densely}, a DCNN with 121 layers pretrained on ImageNet \cite{russakovsky2015imagenet} dataset, is adopted among existing models for converting a chest X-ray image into a feature vector with 1024 dimensions.

\textbf{Phase 2: Image Search} -- In this phase, the query chest X-ray image is first tagged with deep pretrained features. 
Then, the distance between the deep pretrained features of the query chest X-ray image and those of the chest X-ray images in the database are computed. 
The chest X-ray images having the shortest distance with those of the query chest X-ray image are subsequently retrieved.
In this study, Euclidean distance, which is the most widely used distance metric in k-NN \cite{hu2016distance}, is used for computing the distance between the deep features of two given chest X-ray images.

\textbf{Phase 3: Classification} -- 
In this phase, the majority voting of the labels of retrieved chest X-ray images is used as a classification decision.
For example, given a query chest X-ray image, the top $K$ most similar chest X-ray images are retrieved. If $\lceil K/2 \rceil$ chest X-ray images are labelled with pneumothorax, the query image is classified as pneumothorax with a class likelihood of $\lceil K/2 \rceil/K$.

%----------------------------------------------
%-----------------------------------------------

\section{Experiments and Results}
%Our results are summarized as follows.
In this section, we describe the experiments that investigate the use of image search to classify pneumothorax among half a million chest X-ray images, with comparison to traditional classifiers such as Random Forest (RF) \cite{liaw2002classification}.
We first describe the datasets collected and preprocesing procedure, then the experiments, followed by the analysis.

\subsection{Data collection}
Three large public datasets of chest X-ray images were collected.
The first is MIMIC-CXR \cite{goldberger2000physiobank,johnson2019mimic}, a large public dataset of 371,920 chest X-rays associated with 227,943 imaging studies.
Only 248,236 frontal chest X-ray images in the training set were used in this study.
The second dataset is CheXpert \cite{irvin2019chexpert}, a  public dataset for chest radiograph interpretation consisting of 224,316 chest radiographs of 65,240 patients.
Only 191,027 frontal chest X-ray images in the training set were used in this study.
The third dataset is ChestX-ray14 \cite{wang2017chestx}, a  public dataset of 112,120 frontal-view X-ray images of 30,805 unique patients.
All chest X-ray images in this dataset were used in this study. 
In total, 551,383 frontal chest X-ray images were used in this study. 
%A total of 34,605 images (6\% of the dataset) were labelled as pneumothorax. 
The labels refer to the entire image; the collapsed lungs are not highlighted in any way.

%------------------------------------------------------
% Dataset 1
\begin{table*}[htbp]
    \centering
    \caption {A summary of chest X-ray images in the Dataset 1 through combination of three public datasets.} 
    \label{tab:dataset1} 
    \begin{tabular}{lcccc}
        \hline
        ~         & MIMIC-CXR \cite{goldberger2000physiobank,johnson2019mimic} & CheXpert \cite{irvin2019chexpert} & ChestX-ray14 \cite{wang2017chestx} & Total\\ \hline
        +ve: Pneumothorax & 11,610     & 17,693 & 5,302 & 34,605 \\   \hline
        -ve: Normal & 82,668     & 16,974 & 60,361 & 160,003  \\   \hline 
        Total & 94,278     & 34,667 & 65,663 & 194,608  \\   \hline
    \end{tabular}
\end{table*}
%-----------------------------------------------------
%-----------------------------------------------------
% Dataset 2
\begin{table*}[htbp]
    \centering
    \caption {A summary of chest X-ray images in the Dataset 2 through combination of three public datasets.} 
    \label{tab:dataset2} 
    \begin{tabular}{lcccc}
        \hline
        ~         & MIMIC-CXR \cite{goldberger2000physiobank,johnson2019mimic} & CheXpert \cite{irvin2019chexpert} & ChestX-ray14 \cite{wang2017chestx} & Total\\ \hline
        +ve: Pneumothorax & 11,610     & 17,693 & 5,302 & 34,605 \\   \hline
        -ve: Non-Pneumothorax & 236,626     & 173,334 & 106,818 & 516,778  \\   \hline 
        Total & 248,236     & 191,027 & 112,120 & 551,383  \\   \hline
    \end{tabular}
\end{table*}
%-----------------------------------------------------

\subsection{Dataset preparation \& preprocessing}
\textbf{Dataset 1} is a dataset composing of 34,605 pneumothorax chest X-ray images and 160,003 normal chest X-ray images.
The pneumothorax images were obtained from the collected frontal chest x-ray images with the label "Pneumothorax" = 1.
They were considered as positive (+ve) class.
The normal images were obtained from the collected frontal chest x-ray images with label the "No Finidng" = 1.
These chest X-ray images were considered as negative (-ve) class.
% (For ChestX-ray14 \cite{wang2017chestx}, no labels mean no finding.)
A summary is provided in Table \ref{tab:dataset1}.
\textbf{Dataset 2} is a dataset composing of 34,605 pneumothorax chest x-ray images and 516,778 non-pneumothorax chest x-ray images.
The pneumothorax image were obtained from the collected frontal chest X-ray images with the label "Pneumothorax" = 1.
They were considered as positive (+ve) class.
The non-pneumothorax images were obtained from the collected frontal chest X-ray images without the label "Pneumothorax" = 1, meaning that they contain cases such as normal, pneumonia, edema, cardiomegaly, pneumonia and more.
They were considered as negative (-ve) class.
A summary is provided in Table \ref{tab:dataset2}.
It should be noted that ChestX-ray14 \cite{wang2017chestx} dataset was raised with a concern that its chest X-ray images with chest tubes were frequently labelled with Pneumothorax  \cite{zhang2019mitigating,zech2018variable}. 
In our experiments, through combining ChestX-ray14 with CheXpert \cite{irvin2019chexpert}, and MIMIC-CXR \cite{goldberger2000physiobank,johnson2019mimic} datasets, this concern was mitigated to address the bias.

\subsection{Implementation and Parameter Setting}
In this study, DenseNet121 \cite{huang2017densely} %and CheXNet \cite{rajpurkar2017chexnet} 
%was used for feature extraction.
was implemented via the deep learning library Keras (\url{http://keras.io/}) v2.2.4 with Tensorflow backend. %\cite{abadi2016tensorflow}. %was adopted.
Its model weights were obtained through the default setting of Keras.
All images were resize to 224$\times$224 before inputting to the network.
%For CheXNet \cite{rajpurkar2017chexnet}, the model weight file was obtained from the link (\url{https://github.com/brucechou1983/CheXNet-Keras}). 
%The parameter setting is described as follows: 
%The number of training epochs and batch size were 10 and 16 respectively.
% Adam optimizer \cite{kingma2014adam} was used in training.
%All other parameters were set default unless further specified.
Also, Random Forest \cite{liaw2002classification} was implemented using the latest version (v0.20.3) of the machine learning library scikit-learn %\cite{pedregosa2011scikit} 
with the parameter `class\_weight' set to `balanced'.
All other parameters were set default unless further specified.
All experiments were run on a computer with 64.0 GB DDR4 RAM, an Intel Core i9-7900X @3.30GHz CPU (10 Cores) and one GTX 1080 graphics card.
%All DNN were trained on the GPU. 
These settings were used in all experiments unless further specified. 
%The source code is available in \url{https://github.com/antoniosehk/tCheXNet}. 

\subsection{Experiment 1}
%The comparison of the prediction performance among 4 classifiers in terms of the area under the ROC curve in training is summarized in Table \ref{tab:training-roc}. 
%We observe that, except in Dataset-0 and Dataset-6, DeepNN obtained a higher ROC than that of DeepNN-VL (\textbf{Two-tailed Wilcoxon Signed-Rank Test p-value=0.0466 \textless 0.05}). 
%Fig. \ref{fig:roc} shows the comparative results on the area under the ROC curve in the testing set.
%We observe that tChexNet obtained a higher ROC score (10\% better) comparing to that obatained by CheXNet \cite{rajpurkar2017chexnet} in the testing set.
%We observe that in all datasets DNN-VL obtained a higher ROC than DNN and DNN-CL, and the baseline algorithms, SVM (RBF) and Decision Tree.
%We conducted a Two-tailed Wilcoxon Signed-Rank Test on the ROC values obtained by DNN and DNN-VL, as well as the ROC values obtained by DNN-CL and DNN-VL.
%Both tests are statistically significant (\textbf{p=0.00512 \textless 0.05}).
%This is a phenomenon of over-fitting, as also shown by Decision Tree which obtained the highest ROC in training but the lowest ROC in testing.
%Similar results evaluated by the area under the PRC in the testing images are observed in Table \ref{tab:testing-prc}.
In this experiment, we studied the performance of image search to classify pneumothorax among pneumothorax and normal chest X-ray images. 
%The application of such search would be the cases when the radiologist has already a suspicion and hence confines the search to pneumothorax and normal cases.
All chest X-ray images were first tagged with deep pretrained features.
A standard 10-fold cross-validation was then adopted.
All chest X-ray images (with deep pretrained features tagged) were divided into 10 sections.
In each fold, one section of chest X-ray images was used as validation set, while the remaining chest X-ray images were used as training set.
The process was repeated 10 times, such that in each fold a different section of chest X-ray images was used as the validation set.
%The detection performance was evaluated on the validation set in each fold, and then were averaged among 10 folds.

For image search, given a chest X-ray image (from the validation set), it was conducted to search in the training set and used the majority voting of the top $K$ retrieved chest X-ray images to classify.
One experiment was conducted with $K$ = 11 and another was with $K$ = 51.
For Random forest (RF), with number of trees (t) setting as 11 and 51 respectively, it was trained on the deep pretraind features of the training set and evaluated on those of the validation set in each fold.

\begin{figure}[htp]
\centering
\includegraphics[width=0.9\textwidth]{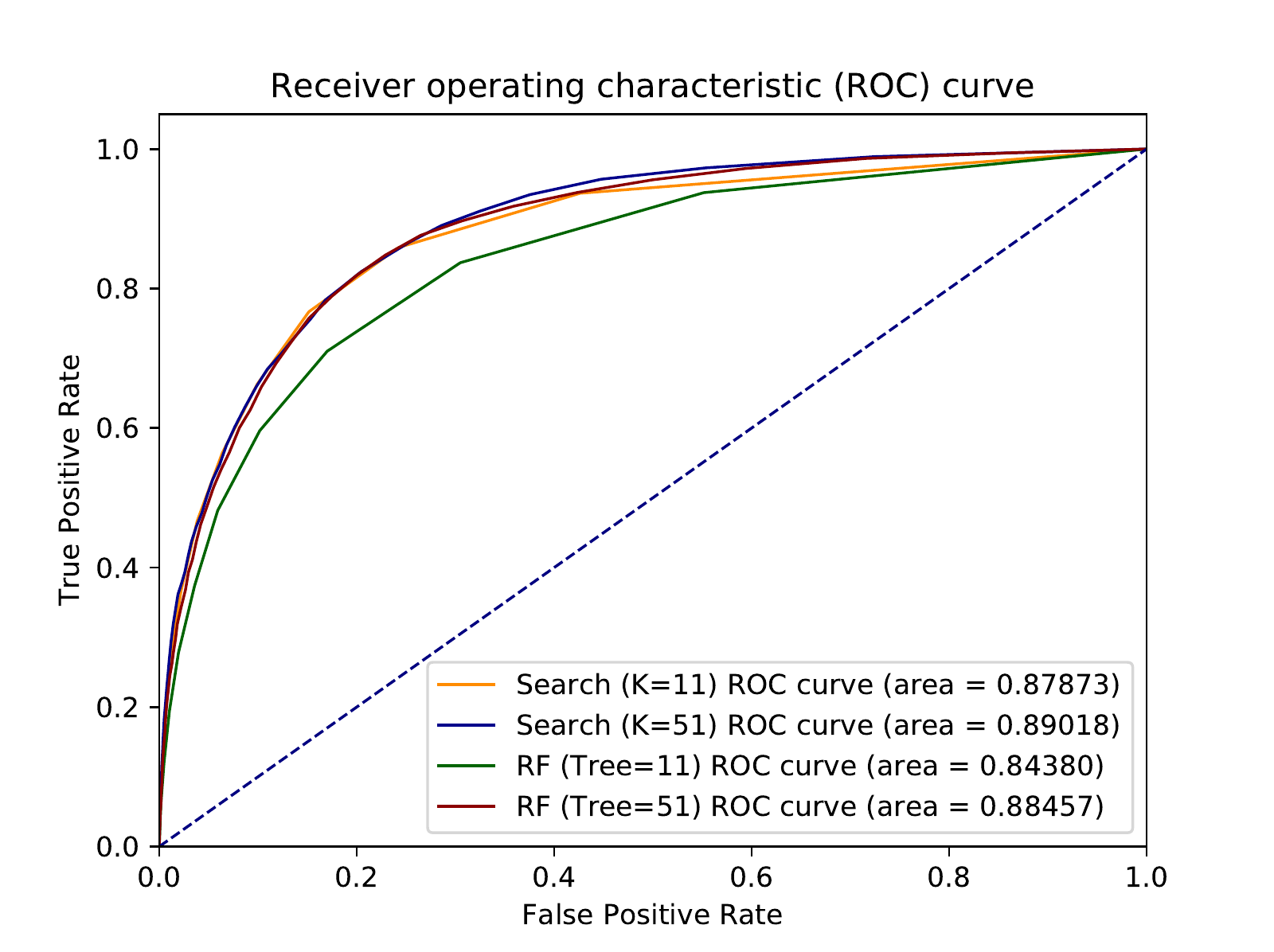}
\caption[Receiver Operating Characteristic (ROC) curve]{%\csentence{
A comparison of the prediction performance in terms of the area under the receiver operating characteristic (ROC) curve on Dataset 1 for Fold-1.
}
\label{fig:roc-dataset1}
\end{figure}

For performance evaluation, following \cite{rajpurkar2017chexnet}, the area under Receiver operating characteristics (ROC) curve was computed for each fold.
A comparison of average area under ROC curve on the results on Dataset 1 under 10-fold cross-validation is summarized in Table \ref{table-px-and-normal}. For completeness and transparency, the result obtained in each fold is demonstrated.
For statistical analysis, two-sample t-test (two-tailed, unequal variance) was conducted on the ROC obtained by RF (t=11) and image search (K=11).
The p-value was \textbf{1.68 E-13 \textless 0.05}, indicating that image search (K=11) obtained a higher ROC, with statistically significance, than that obtained by RF (t=11).
Similarly, the same test was conducted on the ROC obtained by RF (t=51) and image search (K=51).
The p-value was \textbf{5.88 E-3 \textless 0.05}, indicating that image search (K=51) obtained a higher ROC, with statistically significance, than that obtained by RF (t=51). 
% We observed that the use of image search as a classifier obtained better results than classifier.
A ROC curve was provided on fold-1 was provided in Figure \ref{fig:roc-dataset1}.

%-----------------------------------------------------
%-----------------------------------------------------

% Experimental Results 1
\begin{table}[htbp]
  \caption{A comparison of area under ROC curve on Dataset 1 under 10-fold cross-validation between Image Search and Random Forest (RF) trained with deep pretrained features. The p-value of two-sample t-test (two-tailed, unequal variance) between RF (t=11) and Image Search (K=11) is \textbf{1.68 E-13 \textless 0.05}, between RF (t=51) and Image Search (K=11) is \textbf{5.88 E-3 \textless 0.05}.}
  \label{table-px-and-normal}
  \centering
  \begin{tabular}{cccccc}
    \hline
    %\multicolumn{2}{c}{Part}                   \\
    %\cmidrule(r){1-2}
    Fold    & RF (t=11) & Image Search (K=11) & RF (t=51) & Image Search (K=51)\\
    \hline
   1&0.84380&\textbf{0.87873}&0.88457&\textbf{0.89018}\\
   2&0.84906&\textbf{0.87918}&0.88690&\textbf{0.89245}\\
   3&0.84611&\textbf{0.88000}&0.88744&\textbf{0.89214}\\
   4&0.84652&\textbf{0.88118}&0.88788&\textbf{0.89222}\\
   5&0.84658&\textbf{0.88156}&0.88760&\textbf{0.89399}\\
   6&0.84911&\textbf{0.87950}&0.88709&\textbf{0.89192}\\
   7&0.84627&\textbf{0.87220}&0.88464&\textbf{0.88807}\\
   8&0.84210&\textbf{0.87121}&0.87630&\textbf{0.88449}\\
   9&0.85349&\textbf{0.87956}&0.89008&\textbf{0.89351}\\
   10&0.84815&\textbf{0.87461}&0.88572&\textbf{0.88719}\\
    \hline
    mean&0.84703&\textbf{0.87777}&0.88582&\textbf{0.89062}\\
   \hline
  \end{tabular}
\end{table}
%-----------------------------------------------------

\subsection{Experiment 2}
In this experiment, we study the performance of image search to classify if a chest X-ray image has pneumothorax, without any prior knowledge.
%The application of such search would be the cases when we like to deploy image search as a fully automated solution for the pneumothorax diagnosis.
Experimental procedure that was similar to the previous experiment was conducted.
%All chest X-ray images were first divided into 10 sections.
%In each fold, one section of chest X-ray images was used as testing set, while the remaining chest X-ray images were used as training set.
%The above process was repeated 10 times, such that in each fold a different section of chest X-ray images was used as the testing set.
%The detection performance was evaluated on the validation set in each fold, and then were averaged among 10 folds.
%For image search, given a chest X-ray image from the testing set, it was conducted to search among the training set and used the majority voting of the top $K$ retrieved chest X-ray images to classify the input image.
%One was conducted with $K$ = 11 and another was conducted with $K$ = 51.
%The random forest (RF) was trained on the training set and evaluated on the validation set.
%One was trained with 11 trees and the another was trained with 51 trees.
A comparison of area under ROC curve on the results on Dataset 2 is summarized in Table \ref{table-px-and-non-px}.

\begin{figure}[htp]
\centering
\includegraphics[width=0.9\textwidth]{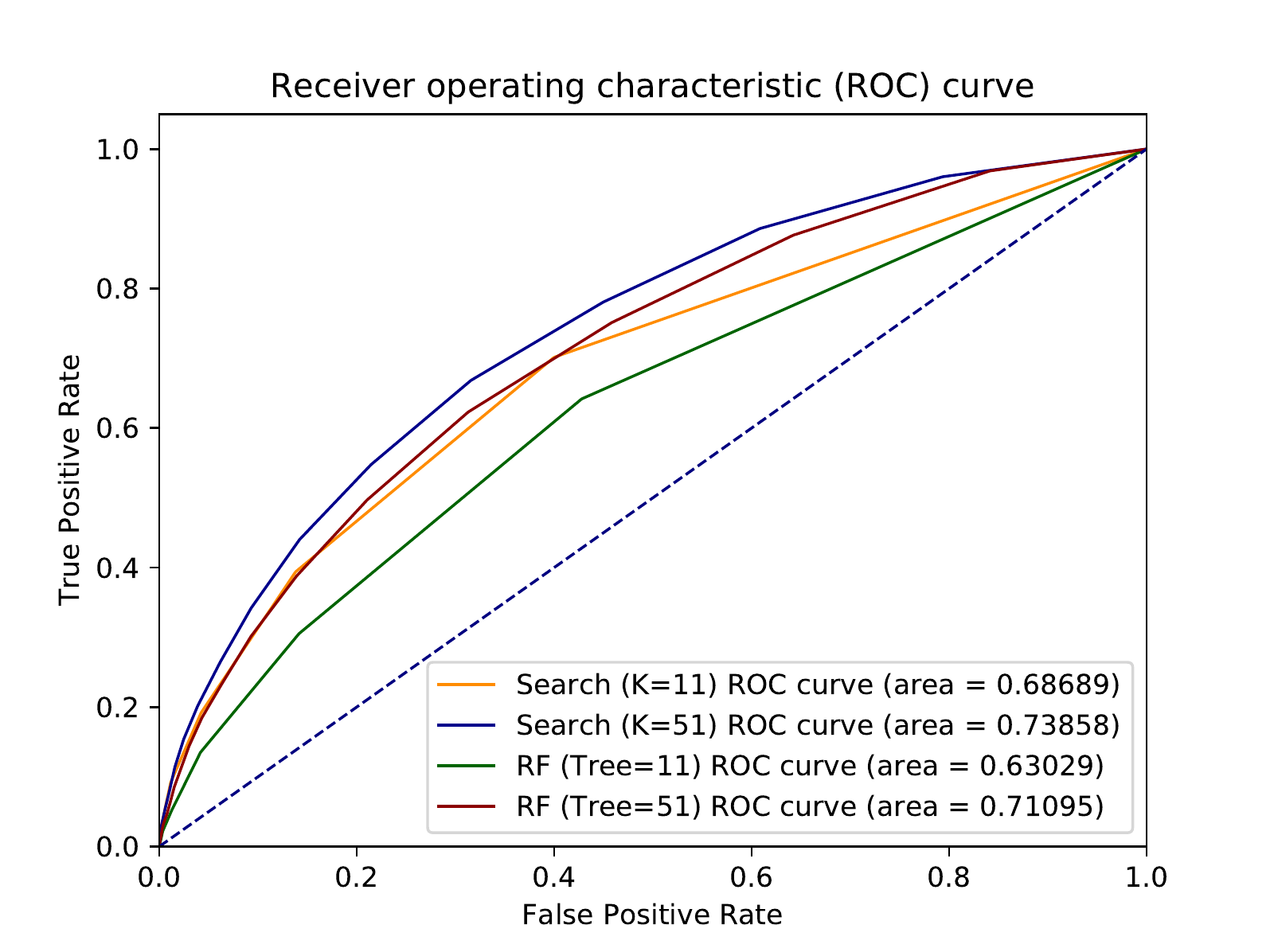}
\caption[Receiver Operating Characteristic (ROC) curve]{%\csentence{
A comparison of the prediction performance in terms of the area under the receiver operating characteristic (ROC) curve on Dataset 2 for Fold-1.
}
\label{fig:roc-dataset2}
\end{figure}

For statistical analysis, two-sample t-test (two-tailed, unequal variance) was conducted on the ROC obtained by RF (t=11) and image search (K=11).
The p-value was \textbf{3.78 E-16 \textless 0.05}, indicating that image search (K=11) obtained a higher ROC, with statistically significance, than that obtained by RF (t=11).
Similarly, the same test was conducted on the ROC obtained by RF (t=51) and image search (K=51).
The p-value was \textbf{5.85 E-12 \textless 0.05}, indicating that image search (K=51) obtained a higher ROC, with statistically significance, than that obtained by RF (t=51). 
%-----------------------------------------------------
% Image search as a classifier obtained better results than RF classifier.
The ROC curve on Fold-1 is provided in Figure \ref{fig:roc-dataset2}.
%-----------------------------------------------------

%-----------------------------------------------------
% Experimental Results 2
\begin{table}[htbp]
  \caption{A comparison of area under ROC curve on Dataset 2 under 10-fold cross-validation between Image Search and Random Forest (RF) trained with deep pretrained features. The p-value of two-sample t-test (two-tailed, unequal variance) between RF (t=11) and Image Search (K=11) is \textbf{3.78 E-16 \textless 0.05}, between RF (t=51) and Image Search (K=51) is \textbf{5.85 E-12 \textless 0.05}.}
  \label{table-px-and-non-px}
  \centering
  \begin{tabular}{cccccc}
    \hline
    %\multicolumn{2}{c}{Part}                   \\
    %\cmidrule(r){1-2}
    Fold    & RF (t=11) & Image Search (K=11) & RF (t=51) & Image Search (K=51)\\
    \hline
    1  & 0.63029 & \textbf{0.68689} & 0.71095 & \textbf{0.73858}\\
    2  & 0.63106 & \textbf{0.69318} & 0.71645 & \textbf{0.74032}\\
    3  & 0.62390 & \textbf{0.69205} & 0.70187 & \textbf{0.73909}\\
    4  & 0.63325 & \textbf{0.69434} & 0.71316 & \textbf{0.74593}\\
    5  & 0.62453 & \textbf{0.68891} & 0.70969 & \textbf{0.74218}\\
    6  & 0.62249 & \textbf{0.68670} & 0.70402 & \textbf{0.73942}\\
    7  & 0.63237 & \textbf{0.69885} & 0.71536 & \textbf{0.75248}\\
    8  & 0.63431 & \textbf{0.68261} & 0.71104 & \textbf{0.73979}\\
    9  & 0.62619 & \textbf{0.69531} & 0.70674 & \textbf{0.74389}\\
    10 & 0.63761 & \textbf{0.68842} & 0.71350 & \textbf{0.74463}\\
    \hline
    mean    & 0.62960 & \textbf{0.69073} & 0.71028 & \textbf{0.74263}\\   
    \hline
  \end{tabular}
\end{table}
%-----------------------------------------------------
\section{Conclusions}
In this study, we explored the use of image search, based on deep pretrained features, in classifying pneumothorax among more than half a million chest X-rays. 
The experiments showed that content-based medical image retrieval system, such as image search, is a potentially viable cost-effective solution. 
To the best of knowledge, it is the first study to demonstrate that deep pretrained features can be used for CBIR of pneumothorax in half a million chest X-ray images.
The setting of image search can be deployed both as a semi-automated and automated solution in the practice of diagnostic radiology. 
Compared with traditional classification, image search results might be clinically more practical as they are supported by the reports and history of evidently diagnosed cases, representing a virtual ``second opinion'' for diagnostic purposes, a factor that may provide more confidence for a reliable diagnosis.

\bibliographystyle{splncs04}
\bibliography{egbib}

\end{document}